\documentclass[aps,preprintnumbers,prl,twocolumn,superscriptaddress]{revtex4}

\usepackage{epsfig,latexsym,cancel,amssymb,amsmath}
\usepackage{graphicx}
\usepackage{epstopdf}
\usepackage{hyperref, graphicx, slashed, xcolor, amsmath}
\usepackage{bm}

\allowdisplaybreaks
\usepackage{color}
\usepackage[normalem]{ulem}
\input{colordvi.tex}

\newcommand{\bx}{{\bf x}}

\newcommand{\bb}{{\bf b}}

\newcommand{\bq}{{\bf q}}
\newcommand{\bp}{{\bf p}}
\newcommand{\bk}{{\bf k}}
\newcommand{\bal}{{\bm{\alpha}}}

\newcommand{\bR}{{\bf R}}
\newcommand{\bG}{{\bf G}}

\newcommand{\bea}{\begin{eqnarray}}
\newcommand{\eea}{\end{eqnarray}}
\newcommand{\nn}{\nonumber\\}

\hypersetup{colorlinks=true,
	breaklinks=true,
	pdfstartview=Fit,
	linkcolor=blue,
	citecolor=blue,
	urlcolor=blue}

\graphicspath{{./figures/}}

\date{September 2024}

\begin{document}
\title{Modulation signals of solar reflected dark matter in crystal-based detectors}
\author{Haipeng An}
	\email{anhp@mail.tsinghua.edu.cn}
	\affiliation{Department of Physics, Tsinghua University, Beijing 100084, China}
	\affiliation{Center for High Energy Physics, Tsinghua University, Beijing 100084, China}
\author{Haoming Nie}
    \email{nhm20@mails.tsinghua.edu.cn}
	\affiliation{Department of Physics, Tsinghua University, Beijing 100084, China}
 
\date{\today}

\begin{abstract}

The scattering of light dark matter (DM) off thermal electrons within the Sun generates a ``fast'' sub-component of the DM flux that can be detected in underground direct detection experiments. This "fast" sub-component has a specific origin—namely, from the Sun. In this study, we demonstrate that in detectors composed of single crystals, like in Bragg scattering, the collision rate and energy deposition are influenced by the angle between the momentum of the incoming DM and the orientations of the crystallographic axes. This results in a directional modulation of the signal. We calculate the magnitude of directional modulations for both germanium and silicon crystals, considering both the contact interaction and light mediator scenarios. Our findings indicate that for the contact interaction case, the daily modulation of the collision rate is approximately 0.1\% of the total, while in the light mediator case, it can reach as high as 30\%. Additionally, our analysis suggests that future ton-scale crystal detectors will be able to explore the freeze-in DM regime with $m_{\rm DM} \sim 0.1 {~\rm MeV}$.

\end{abstract}

\maketitle

\section{Introduction}

The evidence for DM, demonstrated through its gravitational effects across various astrophysical and cosmological scales, remains a key motivation for pursuing physics beyond the Standard Model. As technology in direct detection experiments advances and sensitivity to DM within the galactic halo has approached the threshold for the elastic scattering of neutrino background fluxes—known as the ``neutrino fog''~\cite{ XENON:2024ijk,PandaX:2024muv}, there is growing interest in a range of DM models that extend beyond the conventional WIMP paradigm. In particular, scenarios involving MeV-scale light DM have received increasing attention in recent years~\cite{SENSEI:2023zdf, DAMIC-M:2023gxo, SuperCDMS:2023sql, XENON:2022ltv, LUX:2020car, LZ:2023poo, PandaX:2022xqx, CDEX:2022kcd, CDEX:2022dda}. 

The most efficient way for MeV-scale DM to register energy deposits in a DM detector is by colliding with electrons. However, for most large-scale detectors, the kinetic energy of MeV-scale DM, approximately ($\sim {\cal O}(1)~{\rm eV}$) is insufficient to exceed the detection threshold.
Recent observations have revealed that the light DM flux includes a high-velocity component resulting from scattering with keV-scale electrons in the Sun's core~\cite{An:2017ojc}. This flux of DM particles is commonly referred to as the solar-reflected DM (SRDM). When these SRDM particles interact with electrons in the target material, they can produce an electron recoil signal in the detector. Taking this solar-reflected flux into account, the constraint on the DM-electron scattering cross section for DM mass around $m_{\rm DM} \sim1$ MeV has been established at approximately $\sim 10^{-38}$cm$^2$, assuming the interaction is contact~\cite{An:2017ojc,An:2021qdl,Emken:2021lgc,Emken:2024nox}. If the interaction is mediated by a light mediator, where the interaction resembles Coulombic forces, the bound on the effective charge of DM is restricted to $Q_{\rm eff} < 10^{-9}$ for MeV-scale DM~\cite{An:2021qdl,Emken:2024nox}.

The SRDM flux is not only more energetic than the DM particles present in the galactic halo, but it also originates directionally from the Sun. When the target material in a detector consists of single crystals, the scattering rate and energy deposition can exhibit a complex relationship with the angle between the DM flux and the crystallographic axes. As a result, Earth's motion around the Sun can induce directional modulations in the detection signal~\cite{Emken:2024nox}. The crystal structure determines the period of the modulations. 
Other than modulation signals induced by the earth's motion inside the DM halo~\cite{Stratman:2024sng,Dinmohammadi:2023amy,Budnik:2017sbu,Sassi:2021umf}, all the SRDM are from the Sun, and therefore, as demonstrated in our work, an ${\cal O}(1)$ level directional modulation can be generated. 


In this study, we utilize germanium and silicon crystals as examples to calculate the extent of daily modulation in the direct detection signal attributed to the SRDM flux. In semiconductor crystals, outer shell electrons form directional energy bands; thus, scattering events between SRDM and these electrons may result in angle-dependent signals. For inner shell electrons, the SRDM can create a directional signal through the Bragg scattering. We demonstrate that Bragg scattering of SRDM with inner shell electrons is the primary source of this modulation. 
In this work, we consider both contact and light mediator scenarios, providing a comprehensive understanding of the SRDM induced directional modulation.

\begin{figure}
    \centering
    \includegraphics[width=0.9\linewidth]{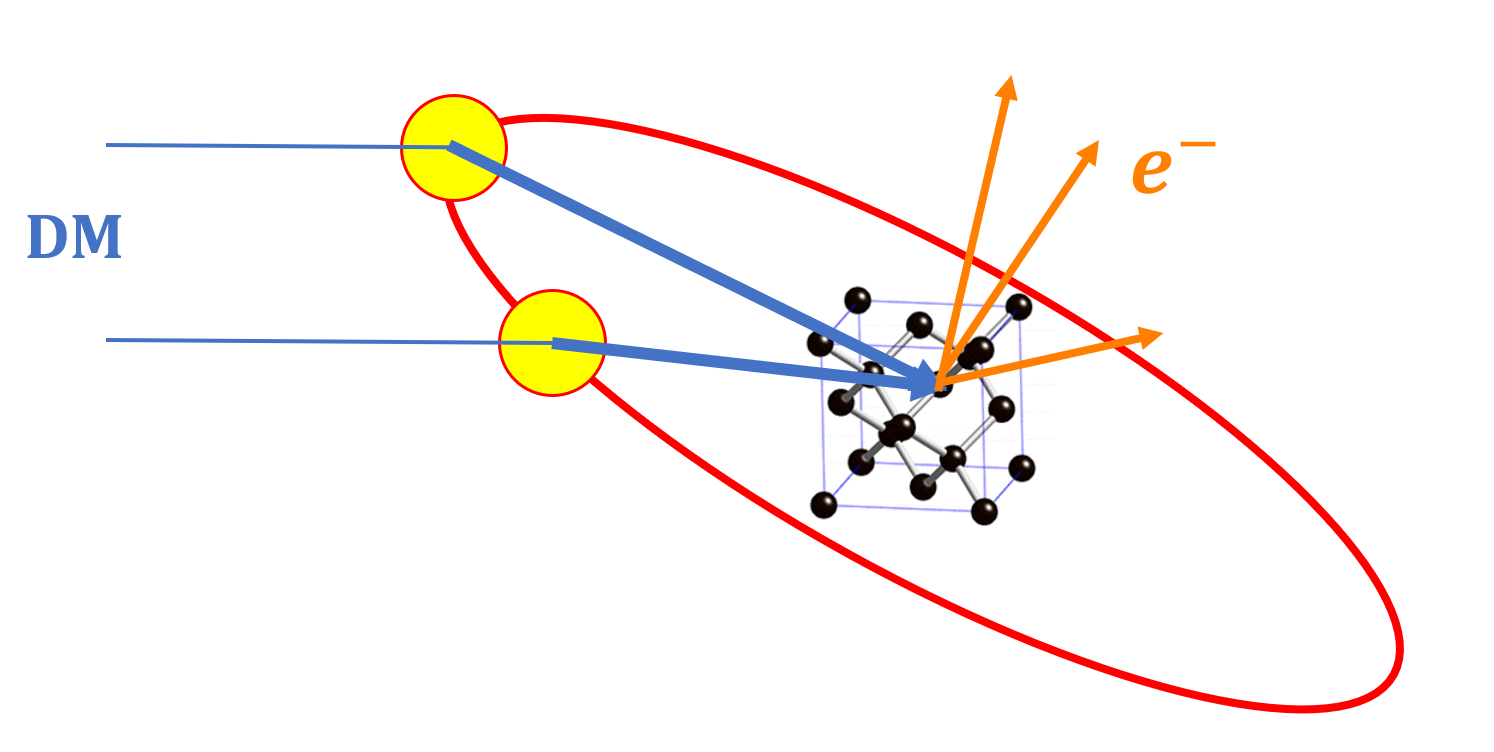}
    \caption{An illustration of direction-dependent scattering in crystal}
    \label{fig:illustration}
\end{figure}

\section{Review of solar reflection DM}

\begin{figure}[htbp]
    \centering
    \includegraphics[width=0.9\linewidth]{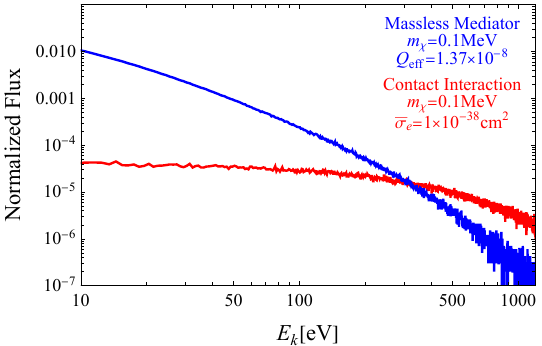}
    \caption{Normalized SRDM fluxes as functions of the DM kinetic energy for both contact interaction case (blue) and light mediator case (red).}
    \label{fig:flux}
\end{figure}

The solar reflected flux of DM has been studied in the literature~\cite{An:2017ojc,An:2021qdl,Emken:2024nox,Emken:2021lgc}. The reflected flux at the Earth's surface can be written as
\begin{equation}    \frac{\mathrm{d}\Phi_{\text{SR}}}{\mathrm{d}E_{\chi}}=\Phi_{\text{halo}}\times\frac{F_{A_\rho}(E_{\chi})A_\rho}{4\pi r_\oplus^2}\ ,
\end{equation}
where $F_{A_\rho}(E_{\chi})$ is the normalized reflected DM flux for the impact disc $A_\rho$ (see~\cite{An:2021qdl} for detailed description of simulation of the reflected flux), $r_\oplus = 1 {\rm AU}$, is the distance between the Sun and the Earth.
Here, $\Phi_{\text{halo}}$ is the halo DM flux. The normalized flux for both the contact interaction and the light mediator cases are shown in Fig.~\ref{fig:flux}, where we can see that the kinetic energies of SRDM are lifted to ${\cal O}(10^2)$ eV scale and the flux in the light mediator case is softer than the contact interaction case. 

\section{The Bragg scattering}

The Bragg scattering matrix element for the DM ($\chi$) to ionize an electron at the inner shell $n$ to an outgoing state with momentum $\bk'$ in a single crystal can be written as~\cite{Thompson:2023jbo}
\bea
{\cal M}_B^{(n)}(\bp,\bq,\bk') = \sum_{{\bf R},i} M^{(n)}_0(\bp,\bq, \bk') e^{i {\bf q} \cdot ({\bm{\alpha}}_i+{\bf R})} \ ,
\eea
where $\bq$ is the momentum transfer, $\bR$ is the crystal lattice vector, $\bal_i$ labels the relative position of atoms in the primitive cell, and $M^{(n)}_0$ is the matrix element for $\chi$ to excite an electron from the inner shell $n$ with momentum transfer $\bq$ in a single atom. 
Then, with the reciprocal lattice vector $\bG$, the Bragg matrix element can be simplified as 
\bea\label{eq:M}
{\cal M}_B^{(n)}(\bp,\bq,\bk')\!=\! M^{(n)}_0(\bp,\bq,\bk')\! S(\bq)\! \frac{(2\pi)^3}{V_{\rm cell}}\!\sum_\bG\! \delta (\bq - \bG)  ,
\eea
where $V_{\rm cell}$ is the volume of a primitive cell, and the factor $S(\bq) = \sum_i e^{i\bq \cdot \bal_i}$. In a crystal, $\bG = h_1 \bb_1 + h_2 \bb_2 + h_3 \bb_3$, where $\bb_{1,2,3}$ are the basis for the reciprocal lattice vectors and $h_{1,2,3}$ are integers. For germanium and silicon crystals $b_1 = b_2 = b_3 =3.790$ and 3.947 keV~\cite{Dillon1957WorkFunctionSO,si_work_func}. 

\begin{figure}
    \centering
    \includegraphics[width=0.65\linewidth]{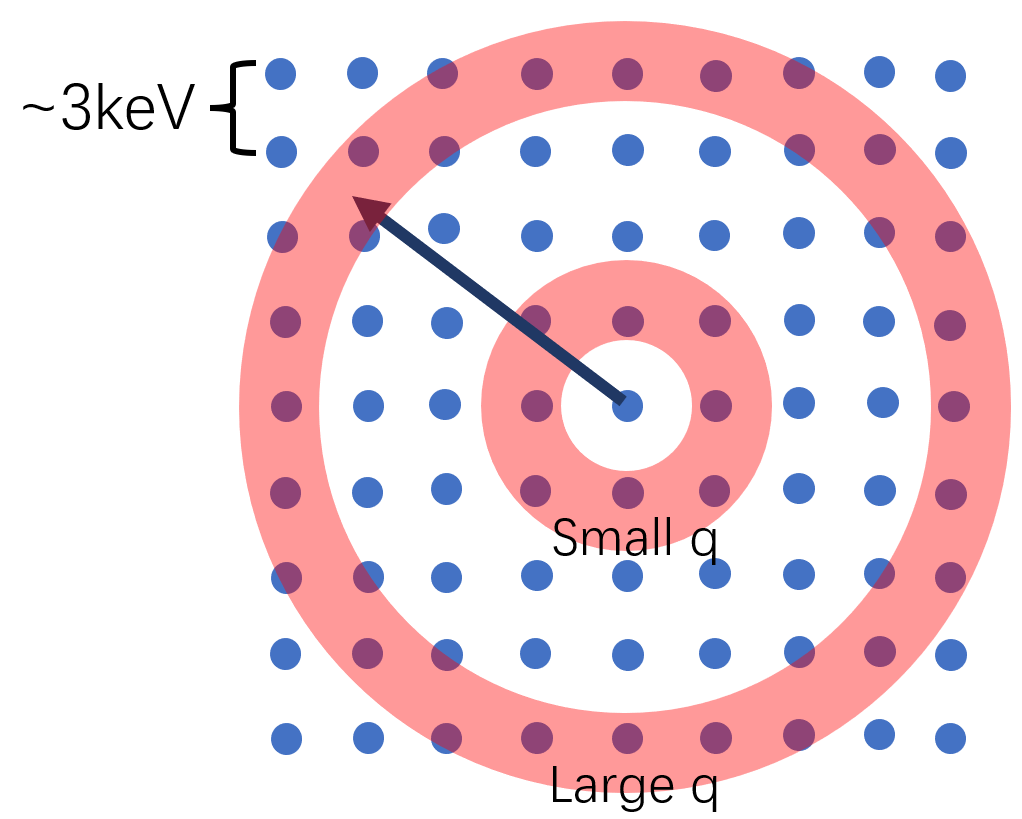}
    \caption{Illustration of directional dependence caused the discrete reciprocal lattice.}
    \label{fig:reciprocal}
\end{figure}

\begin{figure}[htbp]
    \centering
        \includegraphics[width=0.9\linewidth]{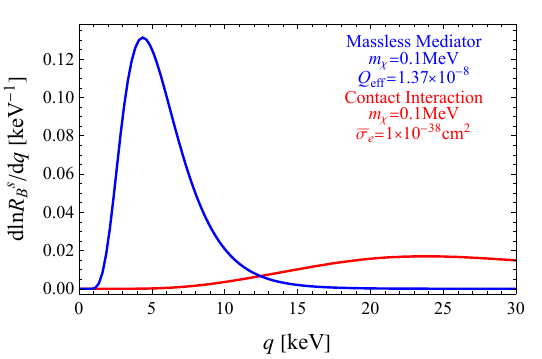}
    \caption{Comparing normalized $d\ln R_B^s / dq$ distributions for individual atoms without Bragg coherence in a germanium detector for both the contact interaction case (red) and the light mediator case (blue).}
    \label{fig:qplot}
\end{figure}

Then, following the standard procedure \cite{Griffin:2021znd}, 
the Bragg scattering rate of SRDM can be written as 
%
\bea\label{eq:rate}
\frac{d R_B}{d E_r}\! =\! N_{\text{cell}}\frac{(2\pi)^3}{V_{\text{cell}}}\!\sum_{\bG\neq \bf 0}\!\int\! d^3 q |S(\bq)|^2 \frac{dR_B^{s}}{d^3 q\,d E_r} \delta^3(\bq - \bG) ,
\eea
where 
\bea\label{eq:Rs}
\frac{dR_B^{s}}{d^3 q\,d E_r}\!=\!\sum_n\!\!\frac{m_\chi^{-1} E_r^{1/2} \Theta(\bq\cdot\hat\bp)}{2^{17/2}\pi^5 m_e^{1/2} (\bq\cdot\hat{\bp})}\!\frac{d\Phi_{\rm SR}}{d E_\chi}\!\left|M^{(n)}_0\!(\bp,\bq,\bk')\right|^2\!
\eea
is the single atom ionization rate with fixed momentum transfer $\bq$. With fixed values of $\bq$, $E_i$, $E_f$ and $\hat\bp$, the energy of the incoming electron is fixed to be 
\bea\label{eq:Echi}
E_\chi = \frac{1}{2}m_\chi\left(\frac{E_f-E_i}{q}+\frac{q}{2m_\chi}\right)^2\left(\frac{\bq}{\bq\cdot\hat{\bp}}\right)^2.
\eea

\begin{figure*}[htbp]
\centering
\includegraphics[height=0.31\linewidth]{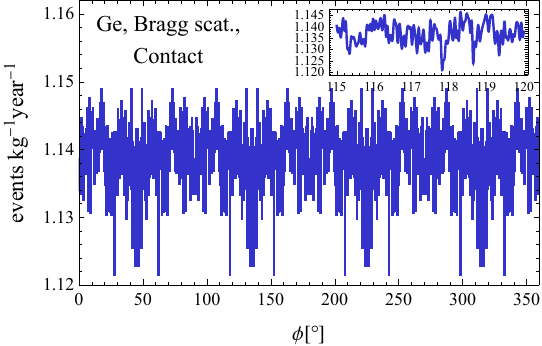}~~~~~
\includegraphics[height=0.31\linewidth]{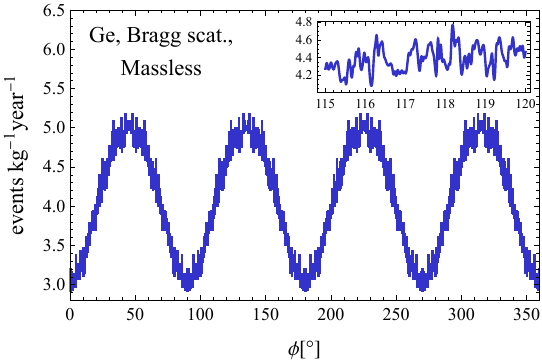}\\
\includegraphics[height=0.31\linewidth]{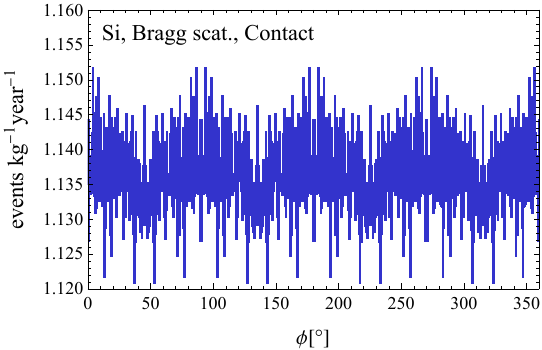}~~~~~
\includegraphics[height=0.31\linewidth]{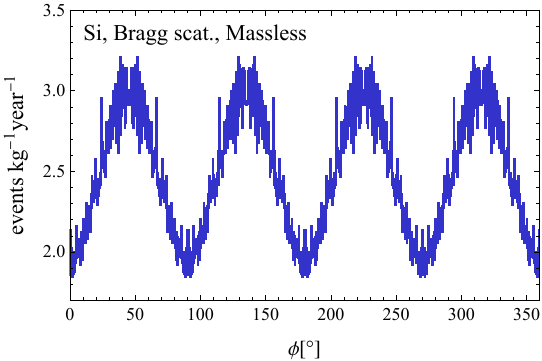}
\caption{The total event rate for SRDM with 0.1 MeV mass scattering off inner shell electrons in germanium crystals (first row) and silicon crystals (second row) with $\bar{\sigma}_e = 1\times 10^{-38}\text{cm}^2$ for the contact interaction case (left) and $Q_{\rm eff} = 1.37\times 10^{-8}$ for the light mediator case (right). The threshold for electron recoil energy is assumed to be 160 eV. 
}
\label{fig:inner_shell_CI}
\end{figure*}

The atomic matrix element squared $|M_0^{(n)}|^2$ can be factorized into three parts,
\bea
|M_0^{(n)}(\bp, \bq,\bk')|^2 \!= \!|F_{\rm DM}(\bp,\bq)|^2 |F_I(q)|^2 |F^{(n)}_e(\bq,\bk')|^2 \ , 
\eea
where $F_{\rm DM}$ and $F_e$ describe the form factors in the DM and atomic parts, and
$F_I$ includes the information of the interaction. The detailed derivations and the explicit forms of the form factors are shown in Appendix A. 
We assume that the interaction between the interactions in both the electron and the DM sides are through monopole of the density distributions. 
Thus, we have $F_{\rm DM} = 1$, and 
\bea
F_e^{(n)}(\bq,\bk') = \int d^3x e^{i \bq\cdot \bx} \psi_{\bk'}^*(\bx)  \psi_b^{(n)}(\bx) \ ,\label{formfactor}
\eea
where $\psi_{\bk'}$ is the wave-function for the outgoing electron with momentum $\bk'$, normalized to $(2\pi)^3 \delta^{(3)}(\bk - \bk')$, and $\psi_b^{(n)}$ is the bound state wave function with quantum numbers labeled by $n$. We use the wavefunctions from \cite{Bunge1993RHF} in the detailed computation.  

In the case of contact interaction, the mediator factor can be written as
\bea
F_I^{(s)}(q^2) = 4m_e m_\chi \frac{1}{\Lambda^2} \ ,
\eea
where $\Lambda$ collects the information of the mediator mass and its couplings to the electron and DM. 
%
%
In the light mediator case, the interaction factor can be parameterized as
\bea\label{eq:FIl}
F_I^{\ell}(q^2) = 4m_e m_\chi\frac{4\pi \alpha_{\rm EM} Q_{\rm eff}}{q^2} ,
\eea
where $\alpha_{\rm EM}$ is the electromagnetic fine structure constant and $Q_{\rm eff}$ is the effective dark charge. For non-relativistic elastic scattering, we have $(q^0)^2 \ll \bq^2$. Thus, we have $F^{\ell}_I = -4m_e m_\chi\cdot4\pi \alpha _{\rm EM} Q_{\rm eff}/\bq^2$. The Debye screening effect in germanium and silicon crystals can be ignored because of the low carrier density. 

\section{Directional dependence in Bragg scattering}

The Dirac delta function $\delta(\bq-\bG)$ in the matrix element (\ref{eq:M}) and the scattering rate (\ref{eq:rate}) reflects Bragg's law, which induces the angular dependence in the scattering rate. As illustrated in Fig.~\ref{fig:reciprocal}, in the $\bq$ integral, in the region that $q$ is comparable to $b_i$, the magnitude of the basis of the reciprocal lattice vector, a significant directional dependence is produced, since $\hat{\bq}$ can only take very few discrete values. However, with $q$ increased, more values of $\hat{\bq}$ becomes available, and therefore the angular dependence becomes suppressed. We can also see that in the region of small $m_\chi$ such that $q \ll b_i$, the scattering rate is heavily suppressed and the crystal becomes "transparent" like glass to visible light because the Bragg condition cannot be satisfied. 
The dominant domain of $q$ that supports the $\bq$ integral in the scattering rate~\ref{eq:rate} is determined by the single atom scattering rate (\ref{eq:Rs}).
In Fig.~\ref{fig:qplot}, we present $d\ln R^s/dq$ with $\hat\bq$ integrated out. We can see that in the case of light mediator interaction the range of $q$ that contributes most to the $dR/dE_r$ is around 3-8 keV. Whereas in the contact interaction case the range lies beyond 10 keV. The reason is in two fold. First, the SRDM spectrum as shown in Fig.~\ref{fig:flux} is much harder in the contact interaction case than in the light mediator case. Second, the explicit dependence of $q^2$ in the denominator of $F^I$ in Eq.~(\ref{eq:FIl}) also makes the peak of $d\ln R^s/dq$ shift to the small $q$ region. Thus, we expect the directional dependence in the light mediator case is more significant than in the contact interaction case.

%
To demonstrate the directional dependence, Fig.~\ref{fig:inner_shell_CI} shows the total scatterings induced by SRDM interactions for $m_\chi = 0.1~{\rm MeV}$ with the inner shells of germanium and silicon crystals. The analysis focuses on recoil energies $E_r > 160 \,\text{eV}$ corresponding to the threshold of the CDEX detector and its future upgrades~\cite{CDEX:2023vvc,Ma:2023yrk}. The scattering rate is plotted as a function of $\phi$, the angle between the Sun-to-Earth direction and the crystallographic x axis of the reciprocal lattice, while the angle between the Sun-to-Earth direction and the z axis is fixed at $90^\circ$. 
The results are presented for both the contact interaction (left) and light mediator (right) cases. For the contact interaction case, the $\chi$-e scattering cross section $\sigma_e$ is set to be $10^{-38}~{\rm cm^2}$, and for the light mediator case, $Q_{\rm eff} = 7\times 10^{-13}$. 
One can observe that the modulation amplitude in the contact interaction case is less than one percent, whereas in the light mediator case, it can reach as high as 30\%. In Fig.~\ref{fig:inner_shell_CI}, 
both panels exhibit angular dependence with periods of approximately $90^{\circ}$. This periodicity arises in the region where $q\sim b_i$, as illustrated in Fig.~\ref{fig:reciprocal}. Additionaly, fluctuations with much faster variations induced are observed, induced by the large $q$ region, as indicated by the spikes in Fig.~\ref{fig:inner_shell_CI}.

\section{SRDM scattering off the valence electrons}

\begin{figure}[htbp]
    \centering
    \includegraphics[width=0.9\linewidth]{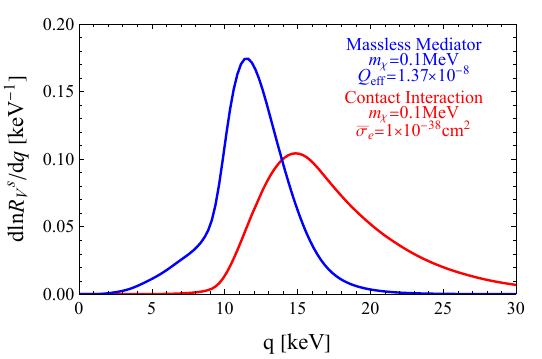}
    \caption{$d\ln R_V^s/dq$ as function of $q$ for valence shell scatterings in germanium for both the contact interaction case (red) and the light mediator case (blue). The parameters are the same as in Fig.~\ref{fig:inner_shell_CI}.}
    \label{fig:qplot_val}
\end{figure}

In a crystal, the valence electrons form plane-wave-like states, known as the Bloch states.
The differential rate of valence shell scattering can be derived as~\cite{Essig:2015cda}
\bea
\frac{dR_V}{dE_r}= N_{\text{cell}} \int d^3 q\frac{d R^s_V}{d E_r d^3 q}, \nn
\eea
where
\bea
\frac{d R^s_V}{d E_r d^3q}&=& \sum_n \int_{\mathbf{BZ}} d^3 k \frac{V_{\text{cell}}^2 m_{\chi}^{-1} E_r^{1/2}\Theta(\bq\cdot\hat{\bp})}{2^{17/2}\pi^7 m_e^{1/2}\bq\cdot\hat{\bp}} \nn
&&\times\frac{d\Phi_{SR}}{dE_\chi}|M^{(n)} (\bq,\bk,\bG)|^2 
\eea
and the relation between $E_\chi$ and $\bq$ is the same as in (\ref{eq:Echi}).
The integration of $\bk$ runs over a full Brillouin zone.
%
In our calculation, we obtain the Bloch state wave functions and calculate the matrix elements using \texttt{Quantum ESPRESSO}~\cite{Giannozzi2017AdvancedCF}, with the lattice parameters from \cite{germanium,Hom1975AccurateLC}. Fig.~\ref{fig:qplot_val} shows $d\ln R_V^s/dq$ for both the contact interaction and light mediator cases for $m_\chi = 0.1~{\rm MeV}$. It is evident that in both cases, the dominant support of the $q$ integral is significantly larger than the size of the base reciprocal vectors, $b_i$. Consequently, the contribution from the valence electrons to the directional signal is expected to be negligible.

\section{Sensitivities of future crystal direct detection experiments}

The periodic nature of directional modulation is governed by the crystal's symmetry. For instance, germanium and silicon crystals (used in CDEX, CDMS, and SENSEI experiments) belong to the Fd-3m space group, which exhibits 2-, 3-, and 4-fold rotational symmetries. These symmetries induce distinct periodic modulations in the signal, with periods of 12 hours, 8 hours, and 6 hours, depending on the crystal orientation relative to the dark matter flux.

To assess the discovery potential of directional modulation in next-generation experiments, we project the sensitivity for CDEX-1T \cite{prospect_cjpl}—a proposed tonne-scale upgrade of CDEX—using a simplified statistical model. The hourly event count, modeled as a Poisson random variable $N_{j,h}$ (where $j$ indexes days and $h$ hours), is defined as
\bea
N_{j,h} = \int^{(h+1){\rm hour}}_{h~{\rm hour}} dt \left[R_{{\rm sig},j}(t) + R_{\rm bg}\right] \ ,
\eea
where $R_{{\rm sig},j}$ is defined as the event rate by integrating $(dR_B/dE_r + dR_V/dE_r)$ over $E_r$ within the detector sensitivity region in the $j$-th day, and $R_{\rm bg}$ denotes the background rate~\cite{CDEX:2023vvc}, assumed to be constant over the observation time.

While solar neutrinos could theoretically produce directional signals in crystal detectors, their interaction cross-section is suppressed by the weak force. At the low recoil energies considered here ($E_r \lesssim 
{\rm keV}$, their contribution is negligible compared to dark matter interactions \cite{Vinyoles_2017, CDEX:2023vvc, solarnuStatus, science.aao0990}.

To maximize the modulation signal,  we align the crystal’s 4-fold symmetry axis with Earth’s rotational axis. In this configuration, the unit vector of incoming SRDM momentum ${\bf p}$ is 
\bea
\hat{\bp}=\left(\sin\theta\cos\left(-\frac{2\pi h}{24}\right),\sin\theta\sin\left(-\frac{2\pi h}{24}\right),\cos\theta\right),
\eea
with $\cos\theta = \sin(2\pi j/365)\sin\theta_\oplus$, where $\theta_\oplus\approx23.5^\circ$ is Earth's axis tilt. A full derivation of the directional dependence is provided in Appendix B.

%
%
%
%
By aligning the crystal’s 4-fold rotational axis with Earth’s rotational axis, the dominant modulation signal exhibits a six-hour periodicity. To isolate this modulated component, we compute the Fourier coefficients:
\bea
S&=&\sum_{j,h} N_{j,h}\sin(2\pi h/6) \equiv 12 D_{\rm tot}S_0 \ ,\nn  C&=&\sum_{j,h} N_{j,h}\cos(2\pi h/6) \equiv 12 D_{\rm tot} C_0  \ ,
\eea
where $D_{\rm tot}$ is the total number of observation days, and $S_0$, $C_0$ represent the amplitudes of the oscillation modes (see Fig.~\ref{fig:inner_shell_CI}). 
Under the Poissonian assumption, the hourly event counts, the hourly event counts
$N_{j,h}$ are uncorrelated random variables. The variances of $S$ and $C$ are thus:
\begin{align}
\text{Var}(S) &= \sum_{j,h} N_{j,h} \sin^2\left(\frac{2\pi h}{6}\right) = 12 \bar{N} D_{\text{tot}}, \\
\text{Var}(C) &= \sum_{j,h} N_{j,h} \cos^2\left(\frac{2\pi h}{6}\right) = 12 \bar{N} D_{\text{tot}},
\end{align}
where $\bar N$ is the average event count per hour.

To quantify sensitivity, we compare the observed statistic $C$ and $S$ to a background-only distribution $C_{\rm BG}$ and $S_{\rm BG}$, generated under the null hypothesis $Q_{\rm eff} = 0$ in the light mediator scenario to set the limit. In our analysis, we choose $h=0$ to be the hour that the signal is in its maximum. Then, $S_0$ vanishes, and all the signals are in the cosine mode.

\begin{figure}
    \centering
    \includegraphics[width=0.96\linewidth]{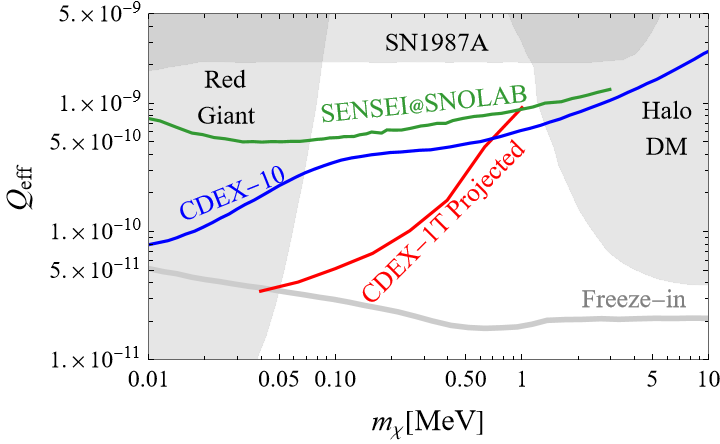}
    \caption{Projected exclusion contour of CDEX-1T with 30 years of exposure from the daily modulation of SRDM in massless mediator case, comparing to CDEX-10 contour without considering modulation \cite{CDEX:2023wfz}. The shaded region on the left comes from red giants cooling \cite{Vogel:2013raa}, the shaded region on the right is a combination of halo DM constraints from \cite{Essig:2017kqs,SENSEI:2020dpa,SENSEI:2023zdf,DAMIC:2019dcn}, the shaded region on the top is the constraint from supernova SN1987A \cite{Chang:2018rso}. The freeze-in benchmark line is also shown for comparison \cite{Essig:2011nj,Essig:2015cda,Chu:2011be,Dvorkin:2019zdi,Emken:2024nox}.}
    \label{fig:scan_plot}
\end{figure}

If $C_{\rm BG}$ lies within the $2\sigma$ confidence interval of $C$, the null hypothesis remains viable, and the corresponding parameter space is excluded from the projected sensitivity. Applying this method to CDEX-1T (Fig.~\ref{fig:scan_plot}, red curve), the projected sensitivity by using the modulation analysis alone is shown by the red curve. This curve means, with such detector, the directional modulation signal can be seen with such $Q_{\rm eff}$. 
As a comparison, the freeze-in benchmark model is also shown. One can see that the sensitivity by the directional modulation analysis alone can already reach the freeze-in benchmark at 0.1 MeV~\cite{Essig:2011nj,Essig:2015cda,Chu:2011be,Dvorkin:2019zdi,Emken:2024nox}.

\section{Summary}

The SRDM flux offers a promising avenue for detecting dark matter particles in the MeV mass range. In this work, we propose leveraging single-crystal detectors to capture directional signatures of SRDM interactions. Specifically, we demonstrate that in scenarios where dark matter couples to electrons via light mediators (as shown in Fig.~\ref{fig:inner_shell_CI}), germanium crystals exhibit a directional modulation effect of up to 30\%. Furthermore, we emphasize that future large-scale detectors employing crystal targets could extend sensitivity to the freeze-in dark matter regime. Directional analysis not only enhances experimental sensitivity in SRDM searches but also aids in distinguishing signal origins and suppressing stationary background contributions, thereby mitigating systematic uncertainties.

\section{Acknowledgement}

We thank Junwu Huang, Hailin Xu and Litao Yang for helpful discussions. This work is supported in part by the National Key R\&D Program of China under Grant No. 2023YFA1607104 and 2021YFC2203100, and the National Science Foundation of China under Grant No. 12475107

\section{Appendix A}

In this appendix, we present the detailed derivation of the formulas that we used to calculate the ionization rate of valence shell electrons in a single-crystal induced by SRDM. The derivation follows the steps in~\cite{Essig:2015cda}.

The ionization matrix element can be written as
\bea
    &&\!\!\!\langle\chi_{\bp-\bq}e_f|H_{\text{int}}|\chi_\bp e_i\rangle\nn
    &&\!\!\!= V\int\frac{d^3 k'}{(2\pi)^3}\psi_f^*(\bk')\langle\chi_{\bp-\bq} e_{\bk'}|H_{\text{int}}\int\frac{d^3 k}{(2\pi)^3}\psi_i(\bk)|\chi_{\bp} e_{\bk}\rangle\nn
    &&\!\!\!\propto\mathcal{M}(\bq)\times V\int\frac{d^3 \bk}{(2\pi)^3}\psi_f^*(\bk+\bq)\psi_i(\bk),
\eea
where $V$ is the volume of the crystal, $e_i$ and $e_f$ denotes the initial and final state of the electron, ${\bf p}$ is the momentum of the incoming SRDM particle, and ${\bf q}$ is the momentum transfer. Therefore the form of scattering rate is different from free-to-free scattering by the replacement
\bea
    (2\pi)^3\delta^{(3)}(\bk-\bq-\bk')|\mathcal{M}|^2\rightarrow|\mathcal{M}|^2 V|F_e(\bq,e_f,e_i)|^2,
\eea
where 
\bea
    F_e(\bq,e_f,e_i)=\int\frac{d^3 \bk}{(2\pi)^3}\psi_f^*(\bk+\bq)\psi_i(\bk).
\eea
This formula is the Fourier transform of Eq.~(\ref{formfactor}) in the main context. 

For inner shell electron ionization the initial wave functions are described by atomic orbitals
\bea
    \psi_{nlm}(\bk) = \chi_{nl}(k)Y^m_l(\hat{\bk}),
\eea
where $\chi_{nl}(k)$ is the Hankel transformation of radial atomic wave function $R_{nl}(r)$, in this study the wave functions are taken from~\cite{Bunge1993RHF}. In principle, we should use the scattering state considering the Coulomb potential for the final state wave function. However, as shown in \cite{Essig:2011nj}, for a full shell it can be approximated by plane waves. The main difference caused by the replacement is that plane waves may have nonzero overlap with atomic orbitals. However, for symmetric atoms with full shells, symmetry requires the summed overlap integral
\bea
\sum_{m}\int d^3 x\ \psi^*_{\bk'}(\bx)\psi_{lm}^{(n)}(\bx)
\eea
must be invariant of any direction of $\bk'$, therefore it must be zero. 
After replacement the square form factor is
\begin{equation}
    |F_e^{(n)}(\bq,\bk')|^2 = \sum_{nl}\frac{2l+1}{k'q}\int_{|k'-q|}^{|k'+q|}kdk|\chi_{nl}(k)|^2.
\end{equation}

Valence shell electrons are delocalized into crystal Bloch states. The Bloch wave function for energy band $n$ is
\bea
\psi^{(n)}_{\bk}(\mathbf{x})=\frac{1}{\sqrt{V}}\sum_{\bG}\tilde{u}_n(\bk+\bG)e^{i(\bk+\bG)\cdot\mathbf{x}},
\eea
Consequently the crystal form factor is
\bea
    &&|F_{\text{crystal}}^{(n)}(\bq,\bk',\bk)|^2\nn
    &&= \frac{(2\pi)^3}{V}\sum_\bG\delta^{3}(\bk-\bk'+\bG+\bq)|\tilde{u}_n(\bk+\bG)|^2,
\eea
where $V$ is the volume of the whole crystal. The above formula depicts the form factor of the whole crystal, for better comparison with the inner shell electron form factor, we multiply it by $N_{\text{cell}}$ and the effective "single electron" form factor used in the main context is
\bea
    &&|F_{e}^{(n)}(\bq,\bk',\bk)|^2\nn
    &&= \frac{(2\pi)^3}{V_{\text{cell}}}\sum_\bG\delta^{3}(\bk-\bk'+\bG+\bq)|\tilde{u}_n(\bk+\bG)|^2.
\eea

\section{Appendix B}

\begin{figure}
    \centering
    \includegraphics[width=0.9\linewidth]{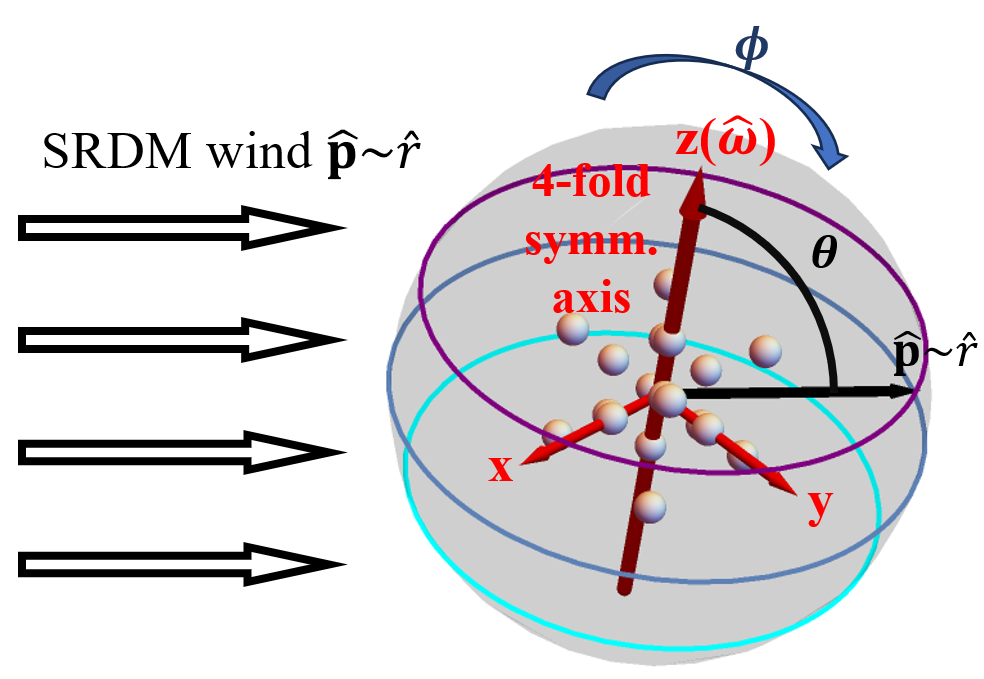}
    \caption{An illustration of $\hat{\bp}$ direction on the celestial sphere. The big red arrow is Earth's rotation axis and at the same time crystallographic z axis. The other red arrows are other crystallographic axes. The Blue curve represents $\hat{\bp}$ in spring and fall equinoxes, the purple curve represents $\hat{\bp}$ in summer solstice, the cyan curve represents $\hat{\bp}$ in winter solstice.}
    \label{fig:CrystalAngle}
\end{figure}

To simplify the analysis, in this work, we assume that the four-fold symmetric axis of the germanium or silicon crystal is placed in alignment to the Earth's rotation axis, as shown in Fig.~\ref{fig:CrystalAngle}. The coordinate system in the analysis of this work is setup such that the $\hat {\bf z}$ direction is along the four-fold symmetric axis of the crystal. The $\hat {\bf x}$ and $\hat {\bf y}$ directions are also shown in Fig.~\ref{fig:CrystalAngle}. Together, they form a Cartesian coordinate system rotating with the Earth. Thus, at the winter solstice and summer solstice, the angle between ${\bf p}$ and ${\bf z}$ are $\pi/2 \mp \theta_\oplus$, respectively. And in the spring and fall equinoxes, ${\bf p}$ is perpendicular to $\hat{\bf z}$. From simple geometric analysis, we can tell that on the $j$-th day after the spring equinox, the angle between ${\bf p}$ and $\hat{\bf z}$ satisfies 
\bea
\cos\theta = \sin(2\pi j/T_y) \sin\theta_\oplus\ ,
\eea
where $T_y$ is the number of days in a year. Then, we can choose the direction of $\hat{\bf x}$ and $\hat{\bf y}$ such that 
\bea
    \hat{\bp}=(\sin\theta\cos\phi,\sin\theta\sin\phi,\cos\theta) \ ,
\eea
where 
\bea
\phi = - 2\pi\times \frac{h}{24} \ .
\eea

\bibliography{ref}

\end{document}